\documentclass[preprint,authoryear,12pt]{elsarticle}

\usepackage[margin=0.80in]{geometry}
\usepackage{float}
\usepackage{adjustbox}
\usepackage{wrapfig}
\usepackage{soul}
\usepackage{footnote}

\usepackage{multirow}
\usepackage{graphics}
\usepackage[dvips]{color}
\usepackage{graphicx}

\usepackage{caption}
\usepackage{subcaption}

\usepackage{amssymb,amstext,amsmath}
\usepackage{float}
\usepackage{booktabs}
\usepackage{array,ragged2e}
\usepackage[table]{xcolor}
\usepackage{multirow}
\usepackage{algorithm}
\usepackage{algpseudocode}
\usepackage{pifont}

\journal{}
\begin{document}
\begin{frontmatter}

\title{Improved Grey System Models for Predicting Traffic Parameters}

\author[MCSaddress]{Gurcan Comert\corref{mycorrespondingauthor}}
\cortext[mycorrespondingauthor]{Corresponding author}
\ead{gurcan.comert@benedict.edu}

\author[MCSaddress]{Negash Begashaw}
\ead{negash.begashaw@benedict.edu}

\author[PEaddress]{Nathan Huynh}
\ead{huynhn@cec.sc.edu}

\address[MCSaddress]{Computer Science, Physics, and Engineering Department, Benedict College, 1600 Harden St., Columbia, USA 29204}
\address[PEaddress]{Department of Civil and Environmental Engineering, University of South Carolina, Columbia, SC 29208}

\begin{abstract}
In transportation applications such as real-time route guidance, ramp metering, congestion pricing and special events traffic management, accurate short-term traffic flow prediction is needed. For this purpose, this paper proposes several novel \textit{online} Grey system models (GM): GM(1,1$|cos(\omega t)$), GM(1,1$|sin(\omega t)$, $cos(\omega t)$), and GM(1,1$|e^{-at}$,$sin(\omega t)$,$cos(\omega t)$).  To evaluate the performance of the proposed models, they are compared against a set of benchmark models: GM(1,1) model, Grey Verhulst models with and without Fourier error corrections, linear time series model, and nonlinear time series model.  The evaluation is performed using loop detector and probe vehicle data from California, Virginia, and Oregon.  Among the benchmark models, the error corrected Grey Verhulst model with Fourier outperformed the GM(1,1) model, linear time series, and non-linear time series models. In turn, the three proposed models, GM(1,1$|cos(\omega t)$), GM(1,1$|sin(\omega t)$,$cos(\omega t)$), and GM(1,1$|e^{-at}$,$sin(\omega t)$,$cos(\omega t)$), outperformed the Grey Verhulst model in prediction by at least 65\%, 16\% and 11\%, in terms of Root Mean Squared Error, and by 82\%, 58\% and 42\%, in terms of Mean Absolute Percentage Error, respectively. It is observed that the proposed Grey system models are more adaptive to location (e.g., perform well for all roadway types) and traffic parameters (e.g., speed, travel time, occupancy, and volume), and they do not require as many data points for training (4 observations are found to be sufficient).
\end{abstract}

\end{frontmatter}



\vspace{-10 pt}
\section{Introduction}
\label{intro}
Traffic congestion has become a way of life for the majority of U.S. commuters in urban areas.  With nearly 90 percent of commuters wanting to get to their destinations at the same time, it is simply a matter of supply not being able to accommodate the demand during peak-hour loads.  Many solutions have been investigated to mitigate traffic congestion, including real-time route guidance, ramp metering, and congestion pricing.  The cornerstone of these solutions is a proactive traffic management system that relies on accurate short-term prediction of traffic parameters, such as speed, travel time, volume, and occupancy (density).  By their nature, these traffic parameters are time series, and therefore, many researchers have attempted to study their behavior using time series models.  However, a shortcoming of time series models is their inability to deal with non-recurring abrupt changes such as those caused by accidents (\cite{2010calling, clements2010forecast, 2008regime}).  This is because traffic flow is nonlinear, stochastic and highly non-stationary.

To overcome the aforementioned limitation with time series models, this paper explores the use of Grey system theory.  As the name implies, Grey (i.e., hazy or fuzzy) systems are those systems without complete information.  Examples of Grey systems include the human body, economy and weather system.  Grey system theory provides methods for resolving latent processes.  As such, it uses discrete or continuous Grey numbers encoded by symbol $\otimes$ to represent the unknown parameters of the system.  Grey models are popular due to  their ability to estimate the behavior of systems using only a small amount of data.  This characteristic makes Grey models suitable for short-term traffic parameter estimation.  

The Grey model GM(1,1) (GM(1,1) stand for  "Grey Model First Order One Variable") is the most widely used and documented in the literature~(\cite{kayacan2010grey}). In \cite{bezuglov2016short}, first author explored the applicability of GM(1,1), Grey Verhulst, and their error corrected versions and showed that Grey models were capable of producing accurate forecasts for un-seasonal or cyclic average traffic speed and travel time data. Authors also noticed slow and excessive forecasts at change points. To improve the performance of the GM(1,1) model for short-term estimation of various traffic parameters, this study proposes to augment the GM(1,1) model with trigonometric functions. Incorporating the oscillating behavior of the sine and cosine functions, or their combinations, these Grey models will be more adept at predicting abrupt changes in traffic flow.  In this paper, the following models are proposed: GM(1,1$|cos(\omega t)$)), GM(1,1$|sin(\omega t)$,$cos(\omega t)$), and GM(1,1$|e^{-at}$,$sin(\omega t)$,$cos(\omega t)$).  To the best of our knowledge, these models have not been studied for short term traffic parameter estimation.  To evaluate the performance of these models, they are compared with a set of benchmark models: GM(1,1) model, Grey Verhulst models with and without Fourier error corrections, seasonal time series model, and nonlinear time series model. The evaluations are done using data from California, Virginia, and Oregon.  The models proposed in this paper outperformed the Grey Verhulst model in prediction by 65\%, 16\% and 11\%, in terms of Root Mean Squared Error, and by 82\%, 58\% and 42\%, in terms of Mean Absolute Percentage Error respectively. It is observed that the proposed Grey system models are more adaptive to location (e.g., perform well for all roadway types) and traffic parameters (e.g., speed, travel time, occupancy, and volume), and they do not require  many data points for training (4 observations are found to be sufficient).

\vspace{-10 pt}
\section{Literature Review}
Grey system theory has been widely applied in different areas of applications for more than 30 years.  Thus, there is much information about its theory and applications in the literature.  Readers are referred to the work of \cite{liu2010grey} for the theory and the work by \cite{Deng1989}.

In transportation related studies, the GM(1,1) model has been used to predict annual average daily traffic (AADT), hourly and monthly volume, traffic accidents, and pavement deterioration. To predict traffic volume, specifically AADT, \cite{zhang2010predicting} combined the GM(1,1) model with a Markov transition matrix. The proposed model performed better in predicting AADTs.  \cite{badhrudeen2016short} compared the performance of the GM(1,1) model against neural network models (NNs).   Using about 100 minutes of observational data to train the GM(1,1) model, the results indicated that the GM(1,1) model outperformed the NNs by 1 to 3\% in mean absolute percent error (MAPE). Similarly, \cite{an2012exploring} compared the performance of the GM(1,1) model with back propagating NN and radial basis NN to predict AADT values.  Their results indicated that the GM(1,1) model outperformed both types of neural network models.  In contrast, \cite{pan2016short}  showed that the GM(1,1) model performed poorly against the NN model in predicting traffic speed (about 6\% on average in MAPE). \cite{yu2015research} examined the GM(1,1) model, the ARIMA (Autoregressive Integrated Moving Average) model, generalized NN regression model and their combinations.  The authors tested different combinations of models with fixed and Elman NN-based weights, a framework similar to forecast encompassing (\cite{clements2010forecast}).  Elman NN-based weighted combined model (i.e., GM(1,1), ARIMA and NN regression combined) showed the highest accuracy in predicting monthly traffic volume. \cite{liu2014highway} combined cubic exponential smoothing with GM(1,1) model.  In their study, they showed that the proposed model provided lower error in predicting the monthly traffic volume.

To predict hourly volume, several researchers (e.g., \cite{shuhua2012city} and \cite{shuhua2010short}) have enhanced the GM(1,1) model by incorporating periodic sine and tangent trigonometric terms. According to \cite{shuhua2012city}, the GM(1,1) model with the sine term included outperformed the GM(1,1) model by 7\% in terms of MAPE in predicting hourly traffic volume. In the study by \cite{shuhua2010short}, the authors optimized the trigonometric function periods for the GM(1,1) model using particle swarm optimization.  The authors did not evaluate the performance of the proposed model. 

Grey models have also been applied to predict traffic accidents and pavement deterioration. Studies in these areas are rather limited. \cite{na2010grey} examined the contributing factors to accidents.  The factors found to have an effect on accident probability, from lowest to highest are: traffic signal timing, driving behavior, weather condition, roadway surface, and visibility. \cite{10331764020150601} applied a variation of the Grey Verhulst model to forecast annual high speed traffic accidents. They reported 9.5\% accuracy in MAPE.  \cite{jiang2005gray} applied the GM(1,1) model to estimate pavement roughness index.  The reported errors in MAPE ranged between 5 and 45\%.

The above review highlights the effectiveness of the GM(1,1) model and its various enhancements.  A particularly promising approach is the incorporation of trigonometric terms into the GM(1,1) model.  This study further contributes to the literature by incorporating a cosine function, a combined sine and cosine function, and an exponential term into the GM(1,1) model for short-term prediction of different traffic parameters.
\vspace{-10 pt}
\section{Methods}
\label{sctGM}
\vspace{-10 pt}
\subsection{GM(1,1) model}
To model time series, Grey System theory provides a family of Grey models, where the first order Grey model is referred to as GM(1,1). The underlying theory of the GM(1,1) model is as follows. Notation was adopted from \cite{kayacan2010grey,bezuglov2016short} and presented here for completeness and initiation to the proposed new Grey models.

Let $X^{(0)}=(x^{(0)}(1),x^{(0)}(2),...,x^{(0)}(n))$ denote a sequence of non-negative observations of a stochastic process and $X^{(1)}=(x^{(1)}(1),x^{(1)}(2),...,x^{(1)}(n))$ be an accumulation sequence of $X^{(0)}$ where 
\begin{equation}
x^{(1)}(k) = \sum_{i=1}^{k}{x^{(0)}(i)}
\label{eq:accumulated_seq}
\end{equation}

The original form of GM(1,1) is defined by the following equation.
\begin{equation}
x^{(0)}(k) + ax^{(1)}(k) = b
\label{eq:original_GM}
\end{equation}

Let $Z^{(1)}=(z^{(1)}(2),z^{(1)}(3),...,z^{(1)}(n))$ be a mean sequence of $X^{(1)}$ where
\begin{equation}
z^{(1)}(k) = \frac{z^{(1)}(k-1)+z^{(1)}(k)}{2}, \forall k = 2,3,\cdots,n
\label{eq:Z_1}
\end{equation}

The basic form of GM(1,1) is given by the following equation.
\begin{equation}
x^{(0)}(k) + az^{(1)}(k) = b
\label{eq:basic_GM}
\end{equation}

If $\hat{a}=(a,b)^T$ and
\begin{eqnarray}
Y = \left[
\begin{array}{c}
x^{(0)}(2) \\
x^{(0)}(3) \\
\vdots \\
x^{(0)}(n) \\
\end{array}
\right],
B = \left[
\begin{array}{cc}
-z^{(1)}(2) & 1 \\
-z^{(1)}(3) & 1 \\
\vdots & \vdots \\
-z^{(1)}(n) & 1 \\
\end{array}
\right].
\label{eq:YandB}
\end{eqnarray}

then, as in~\cite{liu2006grey}, the least squares estimate of the GM(1,1) model is~$\hat{a}=(B^TB)^{-1}B^TY$ and the whitenization equation of the GM(1,1) model is given by,
\begin{equation}
\frac{dx^{(1)}}{dt} + ax^{(1)}(k) = b
\label{eq:whitenization_GM}
\end{equation}

Suppose that $\hat{x}^{(0)}(k)$ and $\hat{x}^{(1)}(k)$ represent the time response sequence (the forecast) and the accumulated time response sequence of GM(1,1) at time $k$ respectively. Then, the latter can be obtained by solving Eq.~(\ref{eq:whitenization_GM}):
\begin{equation}
\hat{x}^{(1)}(k+1)=\left(x^{(0)}(1)-\frac{b}{a}\right)e^{-ak}+\frac{b}{a}, k=1,2,...,n
\label{eq:model_x_1_solution}
\end{equation}

From Eq.~(\ref{eq:accumulated_seq}), the restored values of $\hat{x}^{(0)}(k+1)$ are calculated as $\hat{x}^{(1)}(k+1)-\hat{x}^{(1)}(k)$.  Using Eq.~\ref{eq:model_x_1_solution}, the following prediction equation is obtained.
\begin{equation}
\hat{x}^{(0)}(k+1)=\left(1-e^a\right)\left(x^{(0)}(1)-\frac{b}{a}\right)e^{-ak}, k=1,2,...,n
\label{eq:model_x_0_solution}
\end{equation}

Eq.~(\ref{eq:model_x_0_solution}) is the main forecasting equation that generates values $\forall k=2,3,...,n$.  In the rolling window approach: $x^{(0)}(k+1),x^{(0)}(k+2),...,x^{(0)}(k+w)$, where $w \geq 4$ is the window size (e.g., $w=4$ found to produce very good results \cite{bezuglov2016short}). Forecasts for next data points can be calculated as $\hat{x}^{(0)}(k+w+1), \hat{x}^{(0)}(k+w+2)$.
\vspace{-10 pt}
\subsection{The Grey Verhulst model (GVM)}
\label{sctGVM}
The Grey Verhulst model (GVM) is generally used for more nonlinearly behaving data (\cite{liu2010grey}). Underlying structure of the GVM is given by the following equation \cite{kayacan2010grey,bezuglov2016short}.
%
%
\begin{equation}
x^{(0)}(k)+az^{(1)}(k)=b\left(z^{(1)}(k)\right)^2
\label{eq:verhulst_model}
\end{equation}

The whitenization equation of GVM is:
\begin{equation}
\frac{dx^{(1)}}{dt} + ax^{(1)} = b\left(x^{(1)}\right)^2
\label{eq:verhulst_whitenization}
\end{equation}

Similar to the GM(1,1), the least squares estimate is applied to find~$\hat{a}=(B^TB)^{-1}B^TY$, where
\begin{eqnarray}
Y = \left[
\begin{array}{c}
x^{(0)}(2) \\
x^{(0)}(3) \\
\vdots \\
x^{(0)}(n) \\
\end{array}
\right],
B = \left[
\begin{array}{cc}
-z^{(1)}(2) & z^{(1)}(2)^2 \\
-z^{(1)}(3) & z^{(1)}(3)^2 \\
\vdots & \vdots \\
-z^{(1)}(n) & z^{(1)}(n)^2 \\
\end{array}
\right].
\label{eq:YandB_VerhulstModel}
\end{eqnarray}

The forecasts $\hat{x}^{(0)}(k+1)$ are calculated using Eq.~(\ref{eq:verhulst_x_0_solution}).
\begin{equation}
\hat{x}^{(0)}(k+1)=[\frac{ax^{(0)}(1)\left(a-bx^{(0)}(1)\right)}{bx^{(0)}(1)+\left(a-bx^{(0)}(1)\right)e^{a(k-1)}}][\frac{\left(1-e^a\right)e^{a\left(k-2\right)}}{bx^{(0)}(1)+\left(a-bx^{(0)}(1)\right)e^{a(k-2)}}]
\label{eq:verhulst_x_0_solution}
\end{equation}
%
\vspace{-10 pt}
\subsection{GM(1,1$|sin(\omega t)$) model}
The idea of this proposed model is to enhance prediction of the original GM(1,1) under oscillating behavior. This is accomplished by introducing trigonometric terms on the right hand side (RHS). In order to investigate possible improvements in the model performances, we used a single sine function, a single cosine function, a linear combination of sine and cosine functions, and a linear combination of sine and cosine function multiplied by an exponential term.  GM(1,1$|sin(\omega t)$) model is adopted from \cite{shuhua2012city} and derived via Eq.~\ref{eq:whitenization_5} using the same steps as GM(1,1) and GVM models.
\begin{equation}
\frac{dx^{(1)}}{dt} + ax^{(1)}(k) = b_1 sin(\omega t)+b_2
\label{eq:whitenization_5}
\end{equation}

Model parameters $(\hat{a},\hat{b}_1,\hat{b}_2)^T=(a,b_1,b_2)^T$ are estimated by $(\hat{a},\hat{b}_1,\hat{b}_2)=(B^TB)^{-1}B^TY$ where $Y=[x^{(0)}(2),...,x^{(0)}(n)]^T$ is given above and $B$ is:
\begin{eqnarray}
B = \left[
\begin{array}{ccc}
-z^{(1)}(2) & sin(\omega 2)& 1 \\
-z^{(1)}(3) & sin(\omega 3)&1 \\
\vdots & \vdots \\
-z^{(1)}(n) & sin(\omega n)&1 \\
\end{array}
\right].
\label{eq:YandB5}
\end{eqnarray}

Suppose that $\hat{x}^{(0)}(k)$ and $\hat{x}^{(1)}(k)$ represent the time response sequence (the forecast) and the accumulated time response sequence of GM(1,1$|sin(\omega t)$) at time $k$ respectively. Then, the latter can be obtained by solving the following equation.
\begin{equation}\label{eq:model_5_solution}
\left.\begin{aligned}
x^{(1)}(t+1)&=(x^{(1)}(1)+\frac{b_1\omega}{a^2+\omega^2}-\frac{b}{a})e^{-at}+\\
&\frac{b_1\omega}{a^2+\omega^2}(a sin(\omega t)-\omega cos(\omega t))+\frac{b}{a}
\end{aligned}\right.
\end{equation}
Using the initial condition $x^{(1)}(1)$=$x^{(0)}(1)$ $,\forall k=2,3,...,n$, the restored values of $\hat{x}^{(0)}(k+1)$ in Eq.~(\ref{eq:model_5_solution}) can be calculated as $\hat{x}^{(1)}(k+1)-\hat{x}^{(1)}(k)$. Thus, using Eq.~(\ref{eq:model_5_solution}), the following prediction is obtained.

\begin{equation}\label{eq:model_5}
\left.\begin{aligned}
&\hat{x}^{(0)}(k+1)=(1-e^a)(x^{(0)}(1)+\frac{b_1\omega}{a^2+\omega^2}-\frac{b}{a})e^{-ak}+ \\
&\frac{b_1\omega}{a^2+\omega^2}(a(sin(\omega k)-sin(\omega (k-1)))-\\
&\omega (cos(\omega (k-1))-cos(\omega k)))
\end{aligned}\right.
\end{equation}
%
\vspace{-10 pt}
\subsection{GM(1,1$|cos(\omega t)$) model}
Similar to GM(1,1$|sin(\omega t)$), the model with cosine function is given by the following equation.
\begin{equation}
\frac{dx^{(1)}}{dt} + ax^{(1)}(k) = b_1 cos(\omega t)+b_2
\label{eq:whitenization_6}
\end{equation}

If $(\hat{a},\hat{b}_1,\hat{b}_2)^T=(a,b_1,b_2)^T$ and
\begin{eqnarray}
B = \left[
\begin{array}{ccc}
-z^{(1)}(2) & cos(\omega 2)& 1 \\
-z^{(1)}(3) & cos(\omega 3)&1 \\
\vdots & \vdots \\
-z^{(1)}(n) & cos(\omega n)&1 \\
\end{array}
\right].
\label{eq:YandB6}
\end{eqnarray}

Solution of the differential equation in Eq.~\ref{eq:whitenization_6} is given by Eq.~\ref{eq:model_6_solution}.
\begin{equation}\label{eq:model_6_solution}
\left.\begin{aligned}
&x^{(1)}(t+1)=Ke^{-t}+\\
&(a^{2}b_2 + b_2\omega^2 + a^2 b_1cos(\omega t) + a b_1 \omega sin(\omega t))/(a(a^2 + \omega^2))
\end{aligned}\right.
\end{equation}

where $K$ is obtained from the initial condition  $x^{(1)}(1)$=$x^{(0)}(1)$ and is given by
\begin{equation}\label{eq:model_6_C}
\left.\begin{aligned}
&K=e^{a}[x^{(0)}(1)-(a^{2}b_2 + b_2\omega^2 + \\
&a^2 b_1cos(\omega) + a b_1 \omega sin(\omega))/(a(a^2 + \omega^2))]
\end{aligned}\right.
\end{equation}

Following the same procedure to derive Eq.~(\ref{eq:model_5}) above, using Eq.~(\ref{eq:model_6_solution}), the following prediction equation  is obtained. %
\begin{equation}\label{eq:model_6}
\left.\begin{aligned}
&\hat{x}^{(0)}(k+1)=Ce^{-ak}(1-e^{a})+ \\
&\frac{1}{a^3+a\omega^2}[-a b_1\omega(sin(\omega k)-sin(\omega (k-1)))+\\
&a^2 b_1 (cos(\omega (k-1))-cos(\omega k))]
\end{aligned}\right.
\end{equation}
\vspace{-10 pt}
\subsection{GM(1,1$|sin(\omega t),cos(\omega t)$) model}
The model with a linear combination of Sine and Cosine functions is given by~Eq.~\ref{eq:whitenization_7}.
\begin{equation}
\frac{dx^{(1)}}{dt} + ax^{(1)}(k) = b_1 sin(\omega t)+b_2 cos(\omega t)+b_3
\label{eq:whitenization_7}
\end{equation}

If $(\hat{a},\hat{b}_1,\hat{b}_2,\hat{b}_3)^T=(a,b_1,b_2,b_3)^T$ and
\begin{eqnarray}
B = \left[
\begin{array}{cccc}
-z^{(1)}(2) &sin(\omega 2) &cos(\omega 2)& 1 \\
-z^{(1)}(3) & sin(\omega 3)&cos(\omega 3)&1 \\
\vdots & \vdots \\
-z^{(1)}(n) & sin(\omega n)&cos(\omega n)&1 \\
\end{array}
\right].
\label{eq:YandB7}
\end{eqnarray}

Solution of the differential equation in Eq.~\ref{eq:whitenization_7} is given by,
\begin{equation}\label{eq:model_7_solution}
\left.\begin{aligned}
&x^{(1)}(t+1)=Ke^{-t}+\\
&\frac{(\frac{b_3}{a}+cos(\omega t)(a b_2-b_1\omega)+(ab_1+b_2\omega)sin(\omega t))}{a^2+\omega^2}
\end{aligned}\right.
\end{equation}

where $K$ is obtained from the initial condition $x^{(1)}(1)$=$x^{(0)}(1)$ and is given by
\begin{equation}\label{eq:model_7_C}
\left.\begin{aligned}
&K=e^{a}[x^{(0)}(1)-\\
&\frac{(\frac{b_3}{a}+cos(\omega)(a b_2-b_1\omega)+(ab_1+b_2\omega)sin(\omega))}{a^2+\omega^2}]+\\
\end{aligned}\right.
\end{equation}

Similar to Eq.~(\ref{eq:model_5}) above, Eq.~(\ref{eq:model_7_solution}) can reduce to Eq.~(\ref{eq:model_7}).
\begin{equation}\label{eq:model_7}
\left.\begin{aligned}
&\hat{x}^{(0)}(k+1)=Ce^{-ak}(1-e^a)+\\
&\frac{1}{a^2+\omega^2}(cos(\omega k)-cos(\omega (k-1)))(a b_2-b_1\omega)+\\
&(ab_1+b_2\omega)(sin(\omega k)-sin(\omega (k-1))
\end{aligned}\right.
\end{equation}
\vspace{-10 pt}
\subsection{GM(1,1$|e^{-t},sin(\omega t),cos(\omega t)$) model}
The model with the product of a linear combination of Sine and Cosine functions with an exponential term is given by the following equation.
\begin{equation}
\frac{dx^{(1)}}{dt} + ax^{(1)}(k) = e^{-at}(b_1 sin(\omega t)+b_2 cos(\omega t))+b_3
\label{eq:whitenization_8}
\end{equation}

If $(\hat{a},\hat{b}_3)^T=(a,b_3)^T$ and
\begin{eqnarray}
B = \left[
\begin{array}{cc}
-z^{(1)}(2) & 1 \\
-z^{(1)}(3) & 1 \\
\vdots & \vdots \\
-z^{(1)}(n) & 1 \\
\end{array}
\right].
\label{eq:YandB8}
\end{eqnarray}

Then, if $(\hat{b}_1,\hat{b}_2)^T=(b_1,b_2)^T$
\begin{eqnarray}
B = \left[
\begin{array}{cc}
e^{-2\hat{a}}sin(\omega 2) &e^{-2\hat{a}}cos(\omega 2)\\
e^{-3\hat{a}}sin(\omega 3)&e^{-3\hat{a}}cos(\omega 3) \\
\vdots & \vdots \\
e^{-n\hat{a}}sin(\omega n)&e^{-n\hat{a}}cos(\omega n) \\
\end{array}
\right].
\label{eq:YandB9}
\end{eqnarray}

The solution of the differential equation in Eq.~\ref{eq:whitenization_8} is given by,
\begin{equation}\label{eq:model_8_solution}
\left.\begin{aligned}
&x^{(1)}(t+1)=K e^{-at} - e^{-at}((b_1cos(\omega t)-\\
&b_2sin(\omega t))/\omega - (b_3 e^{at})/a)\\
\end{aligned}\right.
\end{equation}

where $K$ is obtained from the initial condition $x^{(1)}(1)$=$x^{(0)}(1)$ and is given by
\begin{equation}\label{eq:model_8_C}
\left.\begin{aligned}
&K=e^{a}[x^{(0)}(1)+\\
&e^{-a}((b_1cos(\omega) - b_2sin(\omega))/\omega - (b_3e^{-a})/a)]
\end{aligned}\right.
\end{equation}

Following a similar explanation as in Eq.~(\ref{eq:model_5}) above, Eq.~(\ref{eq:model_8_solution}) can be reduced to Eq.~(\ref{eq:model_8}).
\begin{equation}\label{eq:model_8}
\left.\begin{aligned}
&\hat{x}^{(0)}(k+1)=\frac{e^{-ak}}{(1-e^a)}[C+b_1(cos(\omega (k-1))-cos(\omega k))+\\
&b_2(sin(\omega k)-sin(\omega (k-1)))]
\end{aligned}\right.
\end{equation}
\subsection{Error corrections to Grey models}
\label{scterror}
Suppose that $\epsilon^{(0)}$=$\epsilon^{(0)}(1),...,\epsilon^{(0)}(n)$ is the error sequence of $X^{(0)}$, where $\epsilon^{(0)}(k)$= $x^{(0)}(k)-\hat{x}^{(0)}(k)$. These errors can be expressed using Fourier series (\cite{tan1996residual}) as in Eq.~(\ref{eq:fourier_series}). And, expected errors $\hat{\epsilon}^{(0)}(k)$ can be added to $\hat{x}^{(0)}(k)$ (i.e., Grey system model predictions) for all $k$ in order to error correct the predicted values of the signal. For the details of the error correction, readers are referred to \cite{liu2010grey,kayacan2010grey,bezuglov2016short}.

\begin{equation}
\hat{x}_f^{(0)}(k) = \hat{x}^{(0)}(k) + \epsilon^{(0)}(k), k = 2,3,...,n
\label{eq:fourier_correction}
\end{equation}

For simplicity in numerical examples, short notation for Grey models are given in Table~\ref{tab_short}.
\vspace{-10 pt}
\begin{table}[h]
\centering
\caption{Short notation for Grey models}
\label{tab_short}
\scalebox{0.60}{
\begin{tabular}{c c c}
\hline\noalign{\smallskip}
 Model & Short & Error Corrected Model  \\
\noalign{\smallskip}\hline\noalign{\smallskip}
GM(1,1) & GM(1,1) & EFGM\\
GVM & GVM & EFGVM \\
GM(1,1$|sin(\omega t)$) & GM\_S & EFGM\_S\\
GM(1,1$|cos(\omega t)$) & GM\_C & EFGM\_C\\
GM(1,1$|sin(\omega t),cos(\omega t)$) & GM\_SC & EFGM\_SC\\
GM(1,1$|e^{-t},sin(\omega t),cos(\omega t)$) & GM\_ESC & EFGM\_ESC\\
\noalign{\smallskip}\hline\noalign{\smallskip}
\end{tabular}
}
\end{table}
\vspace{-20 pt}
\subsection{Time series benchmark models} 
\label{sctCM}
In addition to the basic Grey system models, linear and nonlinear time series models are used for comparison purposes.  These include autoregressive (Linear) (Eq.(\ref{eqn_linear})), ARIMA (Eq.(\ref{eqn_ARIMAforecast})), SARIMA (Eq.(\ref{eqn_SARIMA})), self-exciting threshold autoregressive model (SETAR) (Eq.(\ref{eqn_setar})) and additive nonlinear autoregressive model (AAR) (Eq.(\ref{eqn_aar})) which is adopted from \cite{di2015package}. These models were also compared in \cite{bezuglov2016short} and \cite{comert2016adaptive}.

For speed forecasts using time series models, AR($\infty$) representation in Eq.~(\ref{eqn_ARinf}) is adopted which provides recursive calculation of coefficients that weighs past observations.
For linear models, examples of fitted models for Eqs.~(\ref{eqn_linear})-(\ref{eqn_SARIMA}) are given in Table~\ref{tab_nlfits}.
\begin{equation}
Z_{t+1}=\phi+\phi_{0}Z_t+\phi_{1}Z_{t-\delta}+...+\phi_{m}Z_{t-(m-1)\delta}+\epsilon_{t+1}
\label{eqn_linear}
\end{equation}
\begin{eqnarray}
\left.\begin{aligned}
\psi(B)(Z_t) = a_t\\
\psi(B)=\frac{\Phi(B)\phi(B)(1-B)^d(1-B)^D}{\theta(B)}
\end{aligned}\right.
\label{eqn_ARinf}
\end{eqnarray}

Using Eq.~(\ref{eqn_ARinf}), one can obtain the forecasting formula for ARIMA($1$,$1$,$2$) as follows.
\begin{eqnarray}
\left.\begin{aligned}
(1-\phi B)(1-B)(Z_t-\mu) = (1-\theta_1 B-\theta_2 B^2)a_t \\
\hat{Z}_{t+1}=Z_t{\sum_{i=1}^{\infty}{B\psi_i}}
\end{aligned}\right.
\label{eqn_ARIMAforecast}
\end{eqnarray}

where $\psi_0 = 1, \psi_1 = 1+\phi-\theta_1, \psi_2 = \psi_1\theta_1-\phi-\theta_2$, and $\psi_i = \psi_{i-1}\theta_1+\theta_2$ for $i>2$. Similarly, SARIMA($1,0,3$)($1,0,0$) forecasts are calculated from Eq.~(\ref{eqn_SARIMA}). ARIMA and SARIMA model parameters (e.g., $\phi$ and $\theta$s) are trained per dataset and traffic parameter and kept constant otherwise.
\begin{eqnarray}
\left.\begin{aligned}
(1-\phi B)(1-\Phi B^{18})(Z_t-\mu) = (1-\theta_1 B-\theta_2 B^2-\theta_3 B^3)a_t \\
\hat{Z}_{t+1}=(1-\sum_{i=1}^{\infty}{\psi_i}\mu)+Z_t{\sum_{i=1}^{\infty}{B\psi_i}} \\
\end{aligned}\right.
\label{eqn_SARIMA}
\end{eqnarray}

where $\psi_0 = 1$, $\psi_1$ = $\phi-\theta_1$, $\psi_2$ = $\psi_1\theta_1-\theta_2$, $\psi_3$ = $\psi_2\theta_1+\psi_2\theta_2-\theta_3$, and $\psi_i$ = $\psi_{i-1}\theta_1+\psi_{i-2}\theta_2+\psi_{i-3}\theta_3$ for $i>3,$ $\psi_i$ = $\psi_{i-1}\theta_1+\psi_{i-2}\theta_2+\psi_{i-3}\theta_3+\Phi$, for $i=18$, $\psi_i$ = $\psi_{i-1}\theta_1+\psi_{i-2}\theta_2+\psi_{i-3}\theta_3-\phi\Phi$, for $i=19$.
\begin{equation}
Z_{t+1}=\begin{cases}\phi_1+\phi_{10}Z_t+\phi_{11}Z_{t-\delta}+...+\phi_{1L}Z_{t-(L-1)\delta}+\epsilon_{t+1},X_t\leq th\\\phi_2+\phi_{20}Z_t+\phi_{21}Z_{t-\delta}+...+\phi_{2H}Z_{t-(H-1)\delta}+\epsilon_{t+1}, X_t> th\end{cases}
\label{eqn_setar}
\end{equation}

where $Z_t$ denotes observed traffic parameters at time $t$, $L=1$ to $5$ and $H=1$ to $5$ low and high regimes, $X_t$ is threshold function (the transition variable), $\delta$ delay of the transition variable, $th$ is the threshold value.
\begin{equation}
Z_{t+1}=\mu+\sum_{j=0}^{m-1}{s_j(Z_{t-(j)\delta})}
\label{eqn_aar}
\end{equation}

In Eq. \ref{eqn_aar}, $s$ represents nonparametric univariate smoothing function that depends on $Z_t$s, $\delta$ is the delay parameter,  $m$ denotes embedding dimension, $D$ is hidden layer of the neural network, and $\beta_i$,$\gamma_{0j}$,$\gamma_{ij}$ represent the weights. Splines from Gaussian family are fitted in the form of $Z_{t+1}\sim\sum_{i=0}^{m-1}{s(Z_{t},..,Z_{t-j})}$.
\section{Data Description}
\label{sctexp}
To assess the performance of the proposed Grey system models to predict various traffic parameters (i.e., speed, travel time, volume, and occupancy) relative to the benchmark models, four different datasets obtained from two different types of collection technologies are used: loop detectors and GPS-based probes. Datasets contain traffic parameters under both normal (without any incidents) and abnormal (with incidents) conditions. Summary of these dataset are given in Table \ref{tab_dataset}.
\begin{figure}[h]
\centering
\begin{subfigure}{.4\columnwidth}
\centering
  \includegraphics[width=0.45\linewidth]{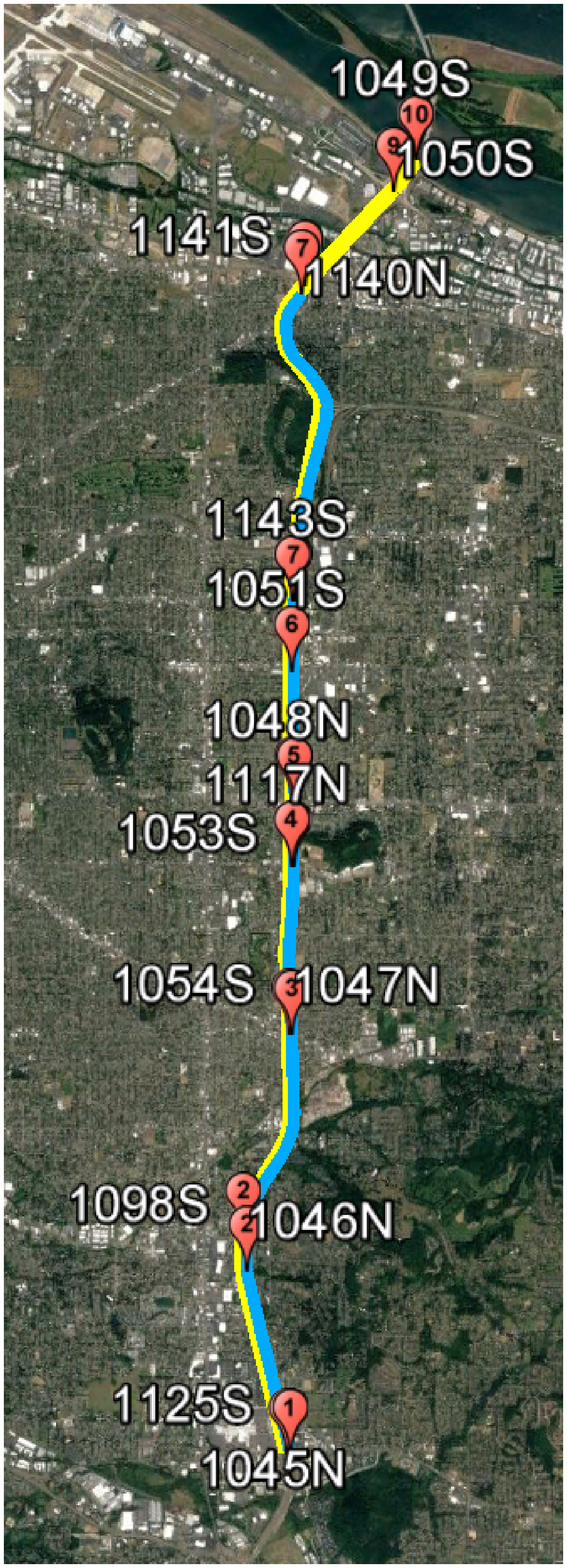}
\caption{Portland Data}
\label{fig_portland}
\end{subfigure}%
\begin{subfigure}{.4\columnwidth}
 \centering
\includegraphics[width=1.25\linewidth, angle=90]{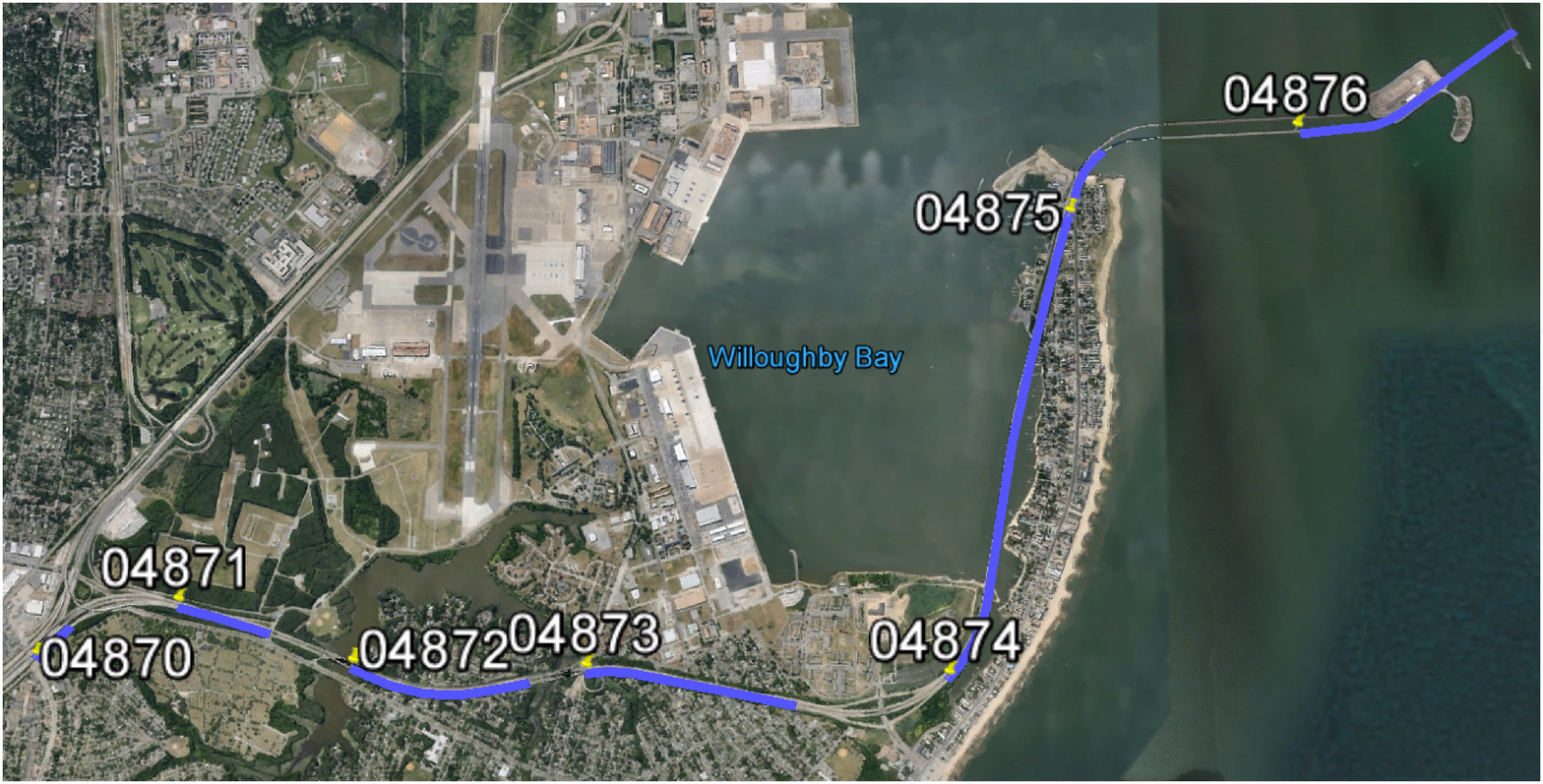}
  \caption{Inrix Data}
  \label{fig_inrix}
\end{subfigure}
\caption{Data Locations}
\label{fig_data}
\end{figure}

The first dataset consists of 1-minute average traffic speeds (S), volume (V), and occupancy (O) collected from loop detectors by the California PATH (Partners for Advanced Transit and Highways) program on I-$880$ for the Freeway Service Patrol Project. The dataset is compiled from $12$ locations for $24$ days in February and March of 1993 between 5:01~AM to 10:00~AM and from 2:01~PM to 7:59~PM.  
\begin{table}[h]
\centering
\caption{Traffic speed, travel time, and volume datasets}
\label{tab_dataset}
\scalebox{0.70}{
\begin{tabular}{c c c c c}
\hline\noalign{\smallskip}
 ID & Date & obs./day & \# of days & Parameter (Agg) \\
\noalign{\smallskip}\hline\noalign{\smallskip}
I-880 Loop Data & Feb16-Mar19 & 560 & 23 & S, V, O (1-min)\\
Inrix Probe Vehicle Data & Jan01-Jan31 & 1440 & 18 & S, TT (1-min) \\
Portland Loop Data & Sep15-Nov15 & 288 & 31 & S, TT, V (5-min)\\
Oregon Loop Data & Sep29-Nov04 & 288 & 6 & S (5-min)\\
\noalign{\smallskip}\hline\noalign{\smallskip}
\end{tabular}
}
\end{table}

The second dataset consists of $20$-second average traffic speeds and travel times (TT) collected from GPS-enabled probe vehicles by INRIX for various freeway segments in Norfolk and Hampton, Virginia, where the longest segment is $1.543$ miles.  This dataset is compiled from $18$ weekdays from January of $2014$. The data is aggregated to $1$-min between $12$:$00$ AM and $12$:$00$ AM (24-hour). Long repeated sections of data are found in this dataset due to lack of probe vehicles especially overnight hours. In order to improve, they were added very low white noise (i.e., $N(0,0.01^2)$) in order to differentiate the missing data. 

The third dataset consists of two months of $20$-second average data (September 15, 2011 through November 15, 2011) collected from dual-loop detectors on I-205, where the Northbound segment is 10.09 miles long and the  Southbound segment is 12.01 miles long.  The collected data include flow, occupancy, and speed as well as calculated travel times.  This dataset is compiled from $11$ different locations (see Fig.~\ref{fig_portland}) 

The fourth dataset has the same traffic parameters as the Portland dataset collected using loop detectors for various locations in the state of Oregon (e.g., I-5, I-405, and I-84) between September 15, 2011 and November 15, 2011. From this dataset, a subset of 6 days between September 29, 2011 and November 4, 2011 was extracted.  This particular dataset contains abnormal fluctuations that were intended to challenge the prediction models.
\begin{table}[h]
	\centering
	\caption{Parameters for time series and Grey system models}
	\label{tab_nlfits}       
	\scalebox{0.70}{
		\begin{tabular}{l l | c}
			\hline\noalign{\smallskip}
			& Model & Parameters \\
			\noalign{\smallskip}\hline\noalign{\smallskip}
			\multirow{3}{*}{SETAR(1,1,1)}& L & $\phi_1=1.215$  $\phi_{10}=0.302$  $\phi_{11}=0.337$ $\phi_{12}=0.221$  \\
			& H &  $\phi_2=2.905$  $\phi_{20}=0.748$  $\phi_{21}=-0.053$ $\phi_{22}=0.196$ \\
			& th & $X_t$=$Z_{t}$, th=12.29 \\
			& proportions & low regime: 69.6 \% \% High regime: 30.4\%  \\ \hline
			\multirow{1}{*}{LINEAR(3)}& & $\phi=0.346$  $\phi_{1}=0.637$  $\phi_{2}=0.146$ $\phi_{3}=0.193$ \\ \hline
			\multirow{1}{*}{ARIMA(1,1,2)}& & $\phi_1=-0.749$  $\theta_{1}=-0.363$ $\theta_{2}=0.402$ \\ \hline
			\multirow{2}{*}{SARIMA(1,0,3)(1,0,0)}& & $\phi_1=0.990$  $\theta_{1}=0.377$  $\theta_{2}=0.121$ $\theta_{3}=-0.047$ \\
			&  &  $\Phi_1=-0.071$ $\mu=11.419$  \\ \hline
			\multirow{1}{*}{GM\_S}& & $\omega=4.30$  \\ \hline
			\multirow{1}{*}{EFGM\_S}& & $\omega=4.30$  \\ \hline
			\multirow{1}{*}{GM\_C}& & $\omega=2.65$  \\ \hline
			\multirow{1}{*}{EFGM\_C}& & $\omega=2.65$  \\ \hline
			\multirow{1}{*}{GM\_SC}& & $\omega=9.30$  \\ \hline
			\multirow{1}{*}{EFGM\_SC}& & $\omega=9.30$  \\ \hline
			\multirow{1}{*}{GM\_ESC}& & $\omega=74.10$  \\ \hline
			\multirow{1}{*}{EFGM\_ESC}& & $\omega=74.10$  \\ \hline
		\end{tabular}		
	}
\end{table}

The benchmark models were trained using the first series of every dataset (i.e., I-880, Inrix, Portland, and Oregon) and every traffic parameter (i.e., speed, travel time, volume, and occupancy). Models were chosen based on the detailed analysis in \cite{bezuglov2016short,comert2016adaptive}. Examples of fitted models using February 16, 1993 I-880 loop occupancy data are presented in Table~\ref{tab_nlfits}. For the Grey system models, the goal is to identify best $\omega$ value for each GM model. The $\omega$ values shown in Table~\ref{tab_nlfits} are optimized only once by grid search on February 16, 1993 I-880 loop speed data and used for other datasets and traffic parameters. The remaining series of each dataset were used to evaluate the models' performance.  
\section{Model Performance and Assessment}
\begin{figure*}[h]
\centering
\begin{subfigure}{0.495\columnwidth}
\centering
  \includegraphics[width=0.9\textwidth]{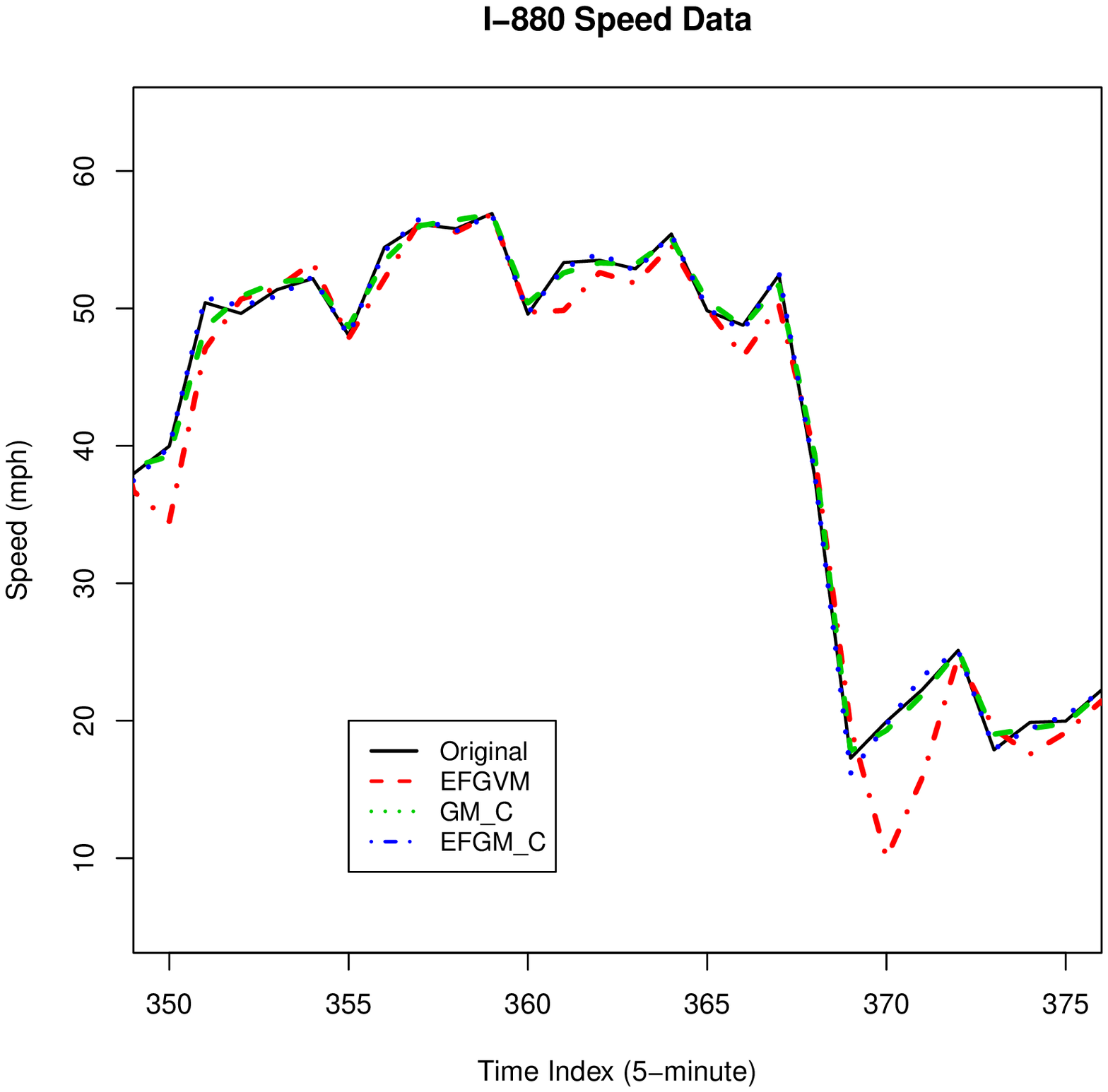}
  \label{fig_fore1a}
\end{subfigure}%
\begin{subfigure}{0.495\columnwidth}
 \centering
\includegraphics[width=0.9\textwidth]{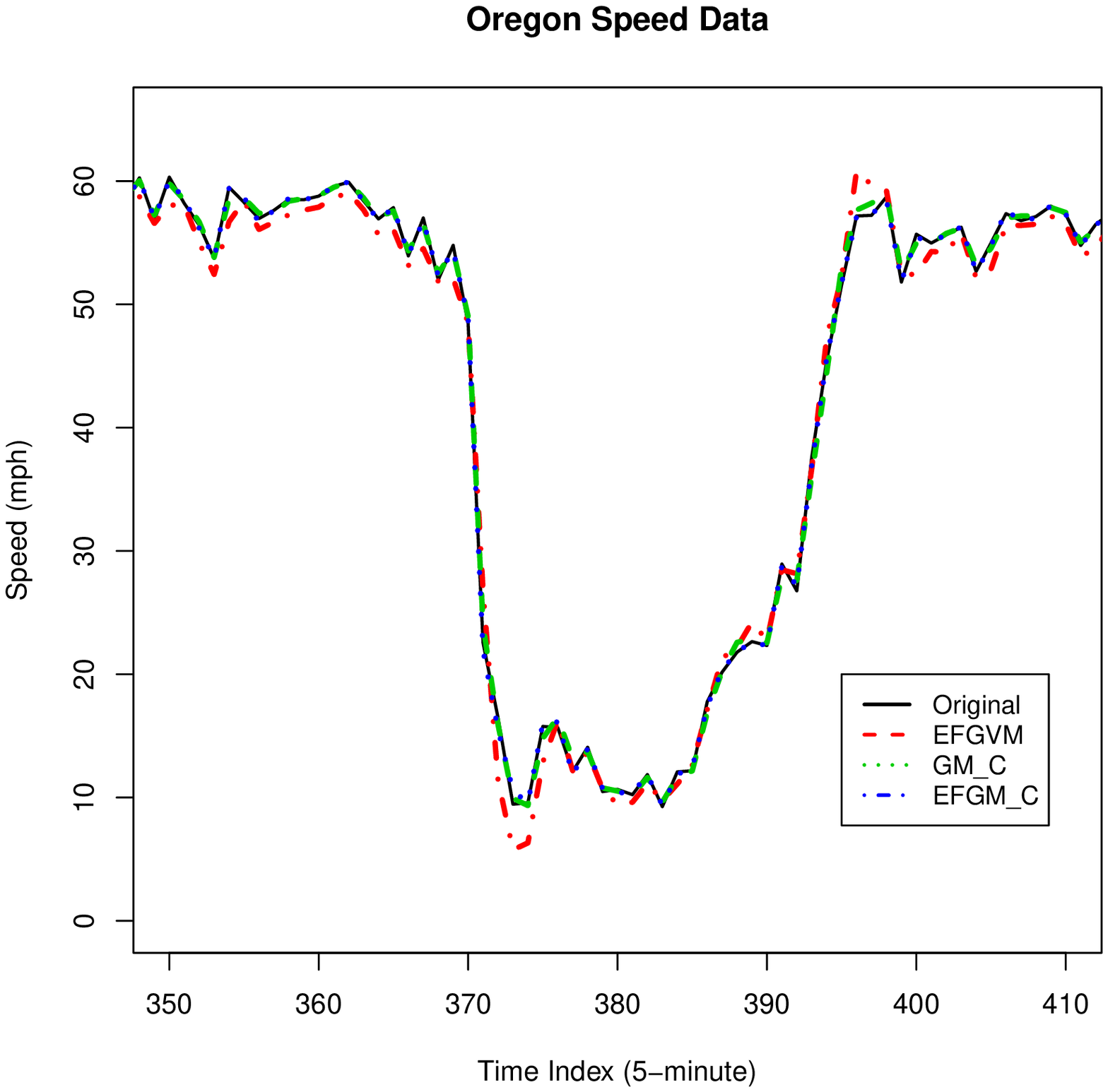}
  \label{fig_fore1b}
\end{subfigure}
\caption{$1$-step speed predictions}
\label{fig_fore1}
\begin{subfigure}{0.495\columnwidth}
\centering
  \includegraphics[width=0.9\textwidth]{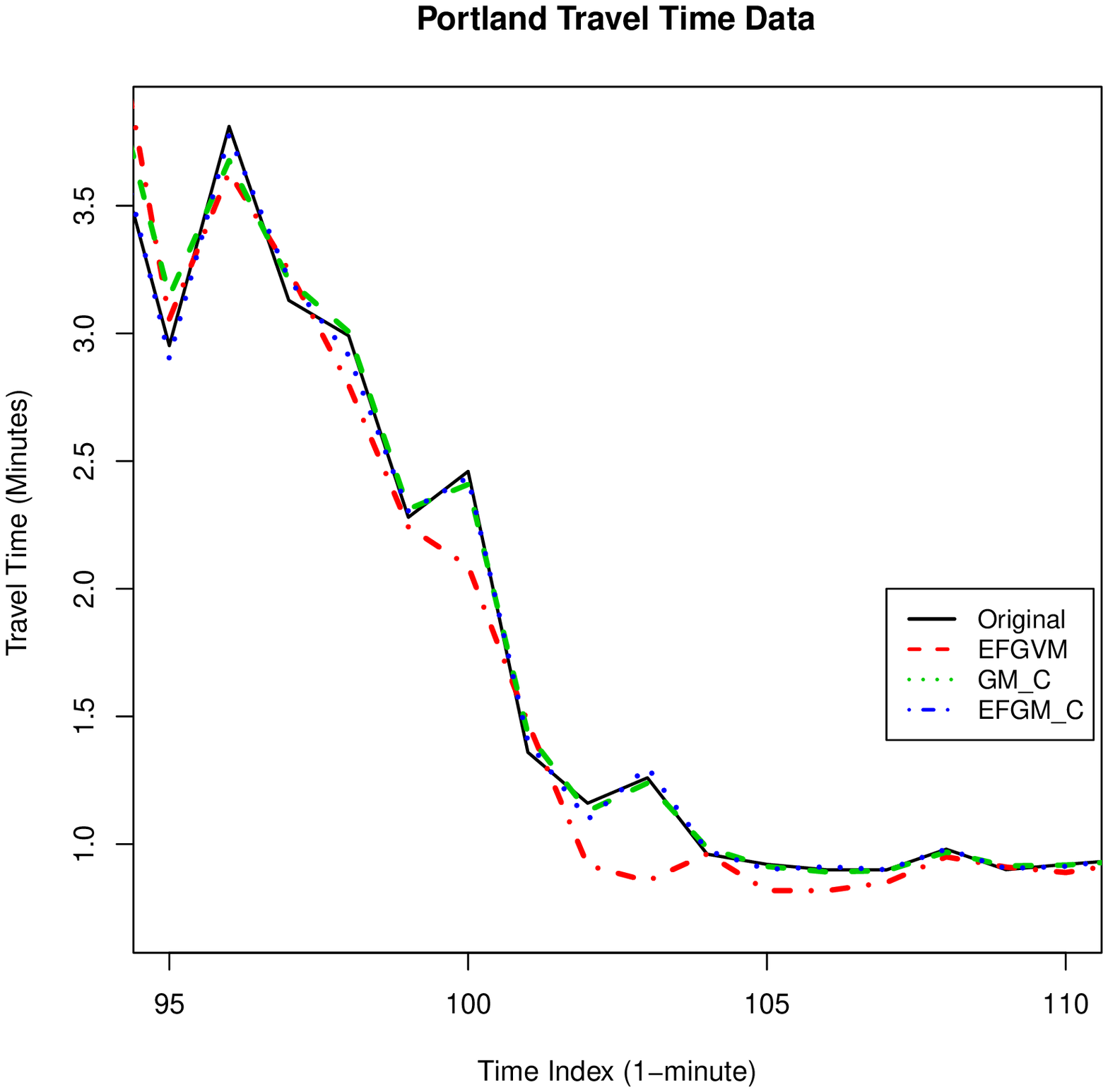}
  \label{fig_fore2a}
\end{subfigure}%
\begin{subfigure}{0.495\columnwidth}
 \centering
\includegraphics[width=0.9\textwidth]{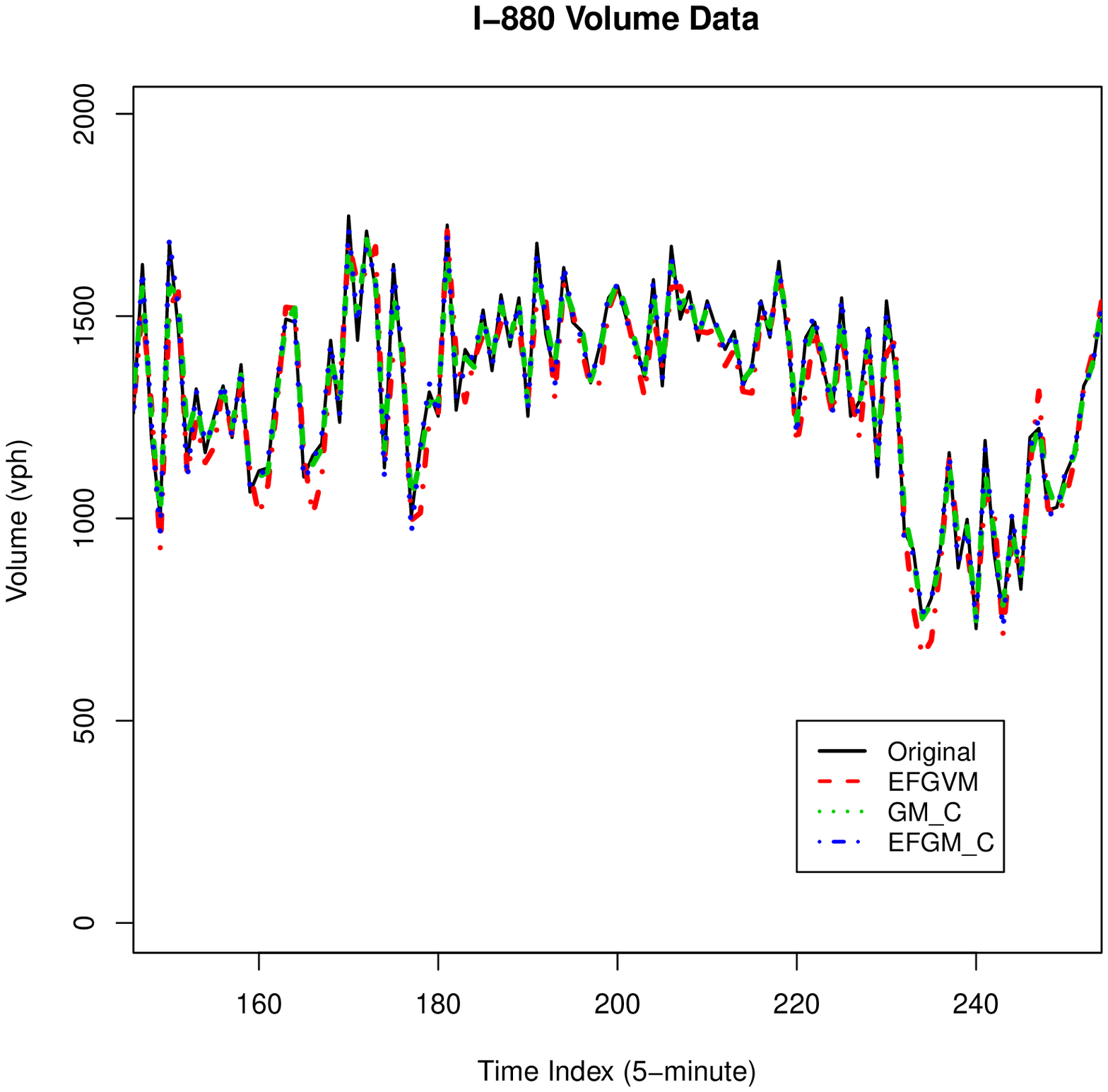}
  \label{fig_fore2b}
\end{subfigure}
\caption{$1$-step travel time predictions}
\label{fig_fore2}
\end{figure*}

Figs.~\ref{fig_fore1}-\ref{fig_fore2} show the prediction performance of the proposed Grey system models compared to actual data under an abnormal condition. To evaluate the performance of the proposed models against the benchmark models, two types of metrics are used: root mean squared error (RMSE) $RMSE_i$ = $\sqrt{\frac{\sum_{k=1}^{N}{\left(\hat{x}_i(k)-x_i(k)\right)^2}}{N}}$ and mean average percentage error (MAPE) $MAPE_i$ = $\frac{1}{N}\sum_{k=1}^{N}{\left|\frac{\hat{x}_i(k)-x_i(k)}{x_i(k)}\right|}100\%$.  These metrics are calculated as follows.

where $\hat{x}_i(k)$ is the predicted and $x_i(k)$ is the observed value at time $k$ for time series $i$. 

Figs.~\ref{fig_fore1}-\ref{fig_fore2} illustrate $1$-step predictions on loop speed and Inrix travel times. Fig.~\ref{fig_fore1} demonstrates the effectiveness of the proposed GM models when the speed quickly increases (time step 220) and decreases (time step 230). The forecasts of best performing GM\_C, EFGM\_C, and EFGVM are shown along with the original values. At speed increases, EFGM overestimates the next value, whereas GM\_C and EFGM\_C provide more conservative estimates, and hence, more accurate forecasts. Since EFGM\_C model is specifically developed to deal with nonlinear sequences, it works better with the traffic speed data which is bound by the speed limits.  Similarly, these models are effective in predicting abrupt changes in travel times as shown in Fig.~\ref{fig_fore2}.
\begin{table}[t!]
\centering
\small
\caption{The accuracy of the models}
\label{tab_accuracy}
\scalebox{0.7}{
\begin{tabular}{r|ll|ll|ll|ll|ll|ll|ll|ll|ll|l}
\hline\noalign{\smallskip}
\multicolumn{1}{c}{Method}& \multicolumn{2}{c}{Portland S} & \multicolumn{2}{c}{Portland V} & \multicolumn{2}{c}{Portland TT} & \multicolumn{2}{c}{Inrix S} & \multicolumn{2}{c}{Inrix TT} & \multicolumn{2}{c}{Oregon S} & \multicolumn{2}{c}{I-880 S} & \multicolumn{2}{c}{I-880 V} & \multicolumn{2}{c}{I-880 O} & \multicolumn{1}{c}{CT} \\
  & RM & MA & RM & MA & RM & MA & RM & MA & RM & MA & RM & MA & RM & MA & RM & MA & RM & MA & (s)\\
   \hline\noalign{\smallskip}
   GM(1,1) & 2.01 & 2.39 & 4.30 & 8.05 & 0.11 & 8.41 & 1.98 & 1.74 & 0.10 & 1.68 & 1.87 & 2.99 & 2.14 & 3.15 & 112.80 & 6.93 & 1.06 & 8.10 & 0.25 \\
    EFGM & 1.86 & 2.44 & 3.76 & 8.28 & 0.10 & 8.42 & 1.51 & 1.37 & 0.07 & 1.31 & 1.73 & 2.82 & 1.91 & 3.08 & 106.27 & 7.14 & 1.02 & 8.29 & 0.61\\
    GVM & 3.27 & 2.33 & 3.77 & 5.46 & 0.08 & 5.35 & 3.02 & 2.26 & 0.10 & 2.18 & 2.89 & 2.40 & 3.12 & 2.61 & 87.99 & 4.41 & 0.80 & 5.16 & 0.25\\
    EFGVM & 1.60 & 2.33 & 2.65 & 5.46 & 0.07 & 5.35 & 1.45 & 2.26 & 0.06 & 2.18 & 1.41 & 2.40 & 1.56 & 2.61 & 68.48 & 4.41 & 0.65 & 5.16& 0.62\\
    GM\_S & 1.53 & 1.71 & 3.39 & 5.75 & 0.08 & 5.70 & 1.62 & 1.50 & 0.08 & 1.41 & 1.45 & 2.20 & 1.61 & 2.24 & 72.52 & 4.60 & 0.77 & 5.53& 0.27\\
    EFGM\_S & 0.91 & 1.13 & 1.82 & 4.16 & 0.05 & 3.87 & 0.78 & 0.66 & 0.04 & 0.61 & 0.86 & 1.30 & 0.95 & 1.48 & 45.00 & 3.04 & 0.45 & 3.62& 0.62\\
    GM\_C & 0.65 & 0.71 & 1.35 & 2.47 & 0.04 & 2.59 & 0.57 & 0.41 & 0.03 & 0.40 & 0.49 & 0.79 & 0.58 & 0.86 & 33.81 & 2.18 & 0.34 & 2.54 & 0.28\\
    EFGM\_C & 0.31 & 0.37 & 0.68 & 1.34 & 0.02 & 1.37 & 0.25 & 0.18 & 0.01 & 0.19 & 0.24 & 0.41 & 0.26 & 0.41 & 17.24 & 1.14 & 0.17 & 1.34& 0.65\\
    GM\_SC & 1.37 & 1.77 & 2.89 & 6.22 & 0.09 & 6.52 & 0.97 & 0.89 & 0.05 & 0.92 & 1.19 & 2.00 & 1.19 & 1.98 & 84.26 & 5.61 & 0.79 & 6.46 & 0.22\\
    EFGM\_SC & 2.00 & 2.67 & 4.25 & 9.00 & 0.12 & 9.68 & 1.50 & 1.39 & 0.07 & 1.41 & 1.85 & 3.10 & 1.91 & 3.13 & 128.26 & 8.48 & 1.19 & 9.69& 0.62\\
    GM\_ESC & 1.36 & 1.84 & 2.77 & 5.93 & 0.08 & 6.25 & 1.13 & 1.32 & 0.05 & 1.17 & 1.26 & 2.08 & 1.39 & 2.29 & 94.91 & 5.50 & 0.76 & 6.18 & 0.21\\
    EFGM\_ESC & 1.98 & 2.75 & 3.75 & 8.03 & 0.10 & 8.43 & 1.66 & 2.28 & 0.07 & 2.08 & 1.82 & 3.02 & 2.00 & 3.37 & 124.90 & 7.43 & 1.02 & 8.35& 0.73 \\
    LINEAR & 3.60 & 6.04 & 6.70 & 16.05 & 0.26 & 23.17 & 6.61 & 11.47 & 0.14 & 2.43 & 3.22 & 5.56 & 3.20 & 5.25 & 168.55 & 11.58 & 1.74 & 14.19& 0.22\\
    SETAR & 3.40 & 5.20 & 6.81 & 15.90 & 0.28 & 24.04 & 7.39 & 12.97 & 0.14 & 2.49 & 3.26 & 5.68 & 3.20 & 5.22 & 170.82 & 11.65 & 1.77 & 14.22 & 0.23\\
    AAR & 4.20 & 6.48 & 6.88 & 16.07 & 0.27 & 24.16 & 3.89 & 5.58 & 0.16 & 2.63 & 3.39 & 6.03 & 3.43 & 5.35 & 169.06 & 11.57 & 1.73 & 13.76 & 2.56\\
    SARIMA & 3.52 & 5.62 & 6.89 & 15.01 & 0.26 & 23.12 & 6.74 & 11.72 & 0.14 & 2.44 & 3.21 & 5.56 & 3.19 & 5.24 & 160.29 & 10.76 & 1.72 & 13.90& 0.56\\
    ARIMA & 3.56 & 5.43 & 7.00 & 14.89 & 0.19 & 14.61 & 2.98 & 3.35 & 0.13 & 2.18 & 3.20 & 5.15 & 3.20 & 5.20 & 160.51 & 10.85 & 1.73 & 13.98& 0.01\\
    \% Imp &59\%	&69\%	&49\%	&55\%	&42\%	&52\%	&61\%	&82\%	&51\%	&82\%	&65\%	&67\%	&63\%	&67\%	&51\%	&51\%	&48\%	&51\% \\
   \hline
\end{tabular}
}
\end{table}

\begin{figure*}[h!]
\centering
\begin{subfigure}{0.48\textwidth}
\centering
  \includegraphics[width=0.81\linewidth]{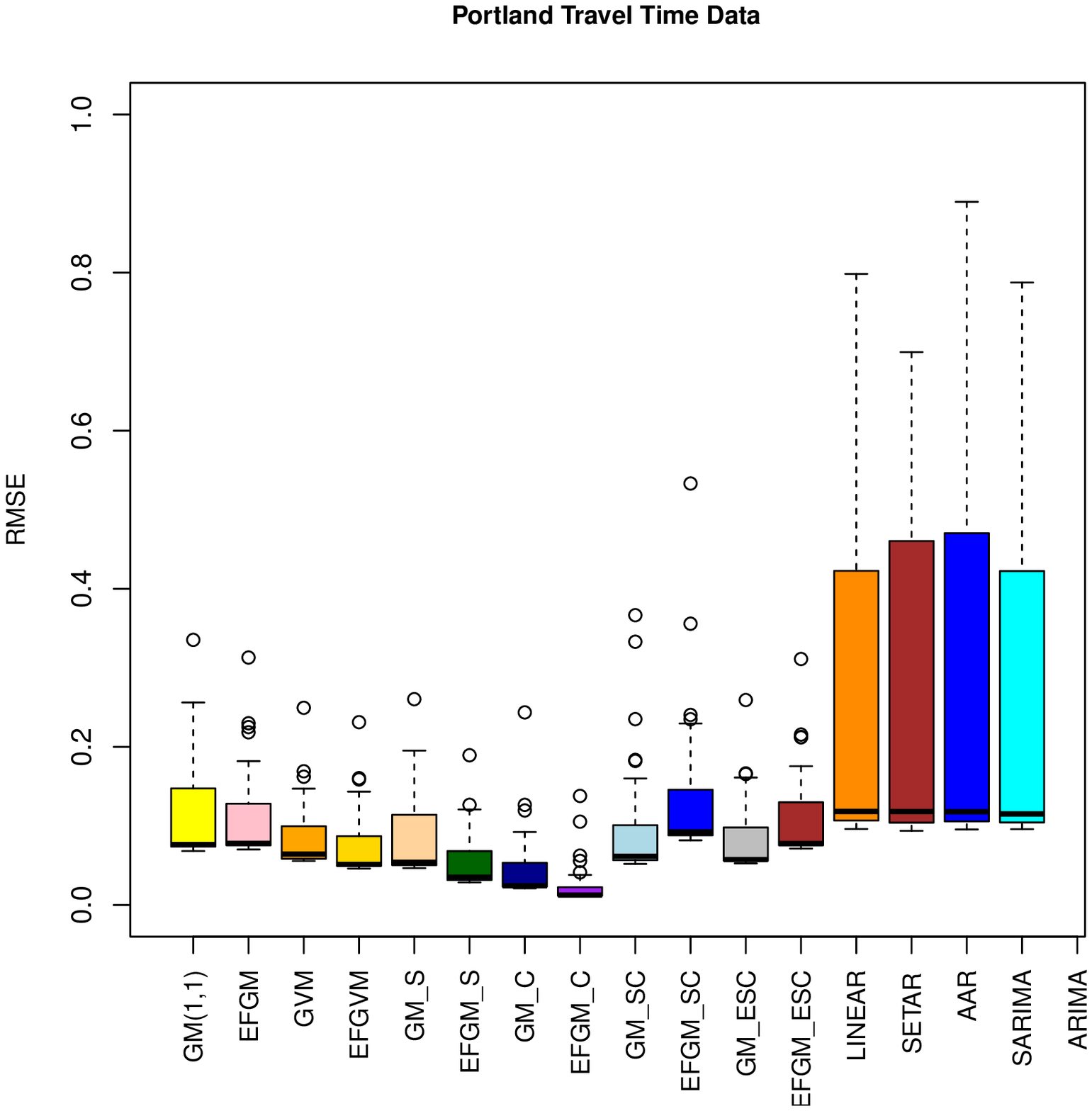}
  \label{fig_boxplot1a}
\end{subfigure}%
\begin{subfigure}{0.47\textwidth}
 \centering
\includegraphics[width=0.81\linewidth]{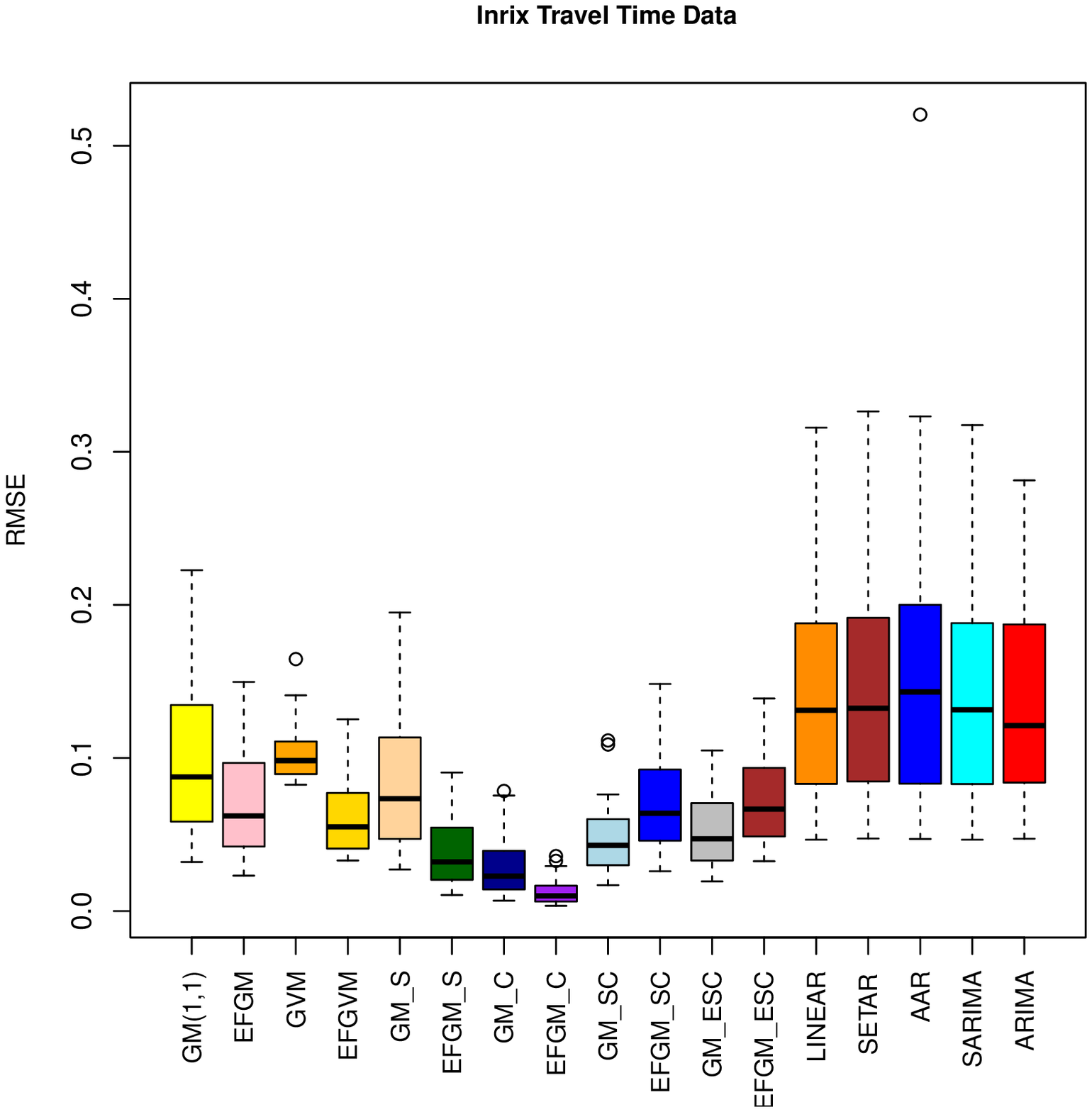}
  \label{fig_boxplot1b}
\end{subfigure}
\caption{RMSEs for $1$-step travel time predictions}
\label{fig_boxplot1}
\begin{subfigure}{0.47\textwidth}
\centering
  \includegraphics[width=0.81\linewidth]{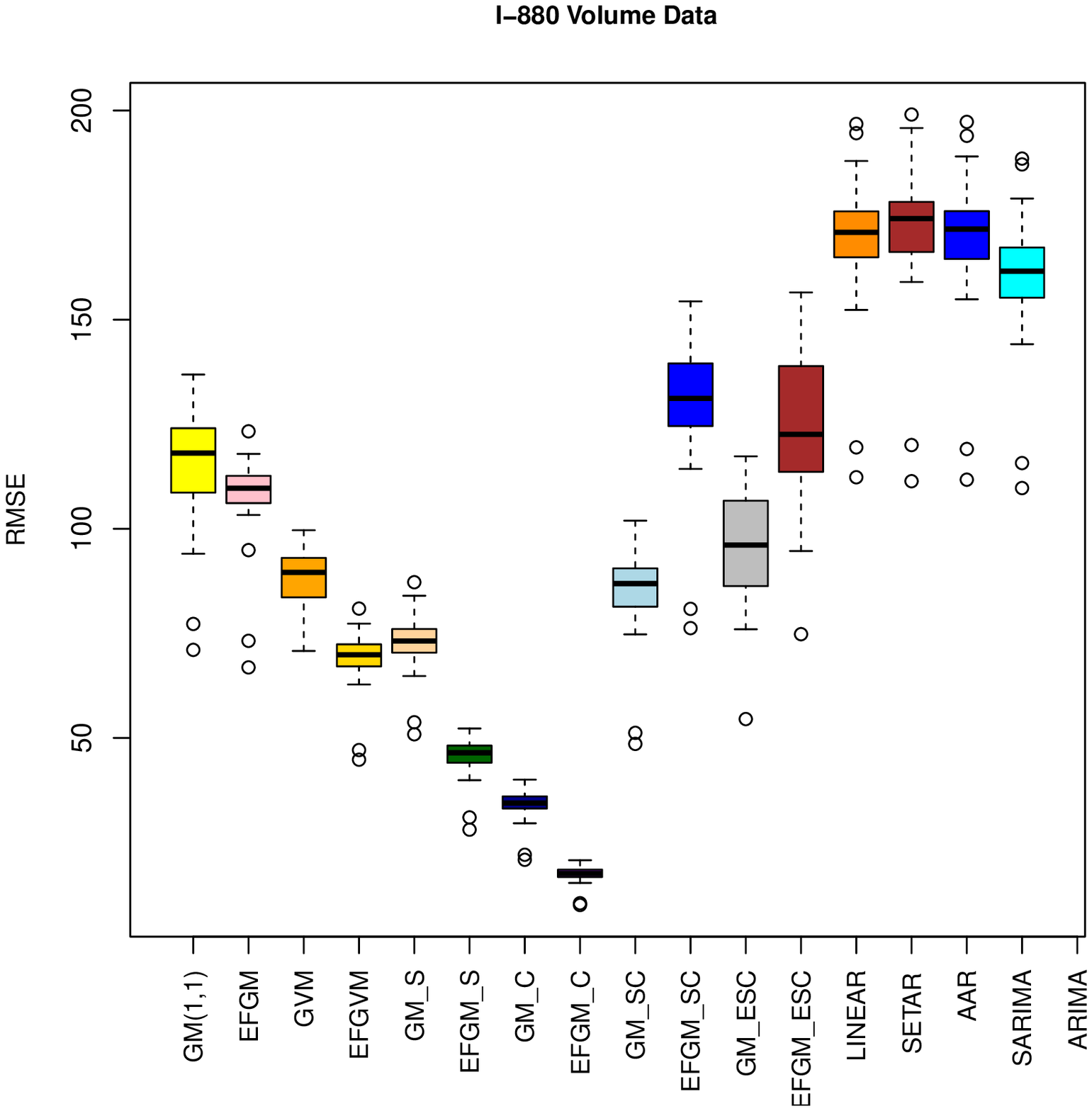}
  \label{fig_rmse}
\end{subfigure}%
\begin{subfigure}{0.47\textwidth}
 \centering
\includegraphics[width=0.81\linewidth]{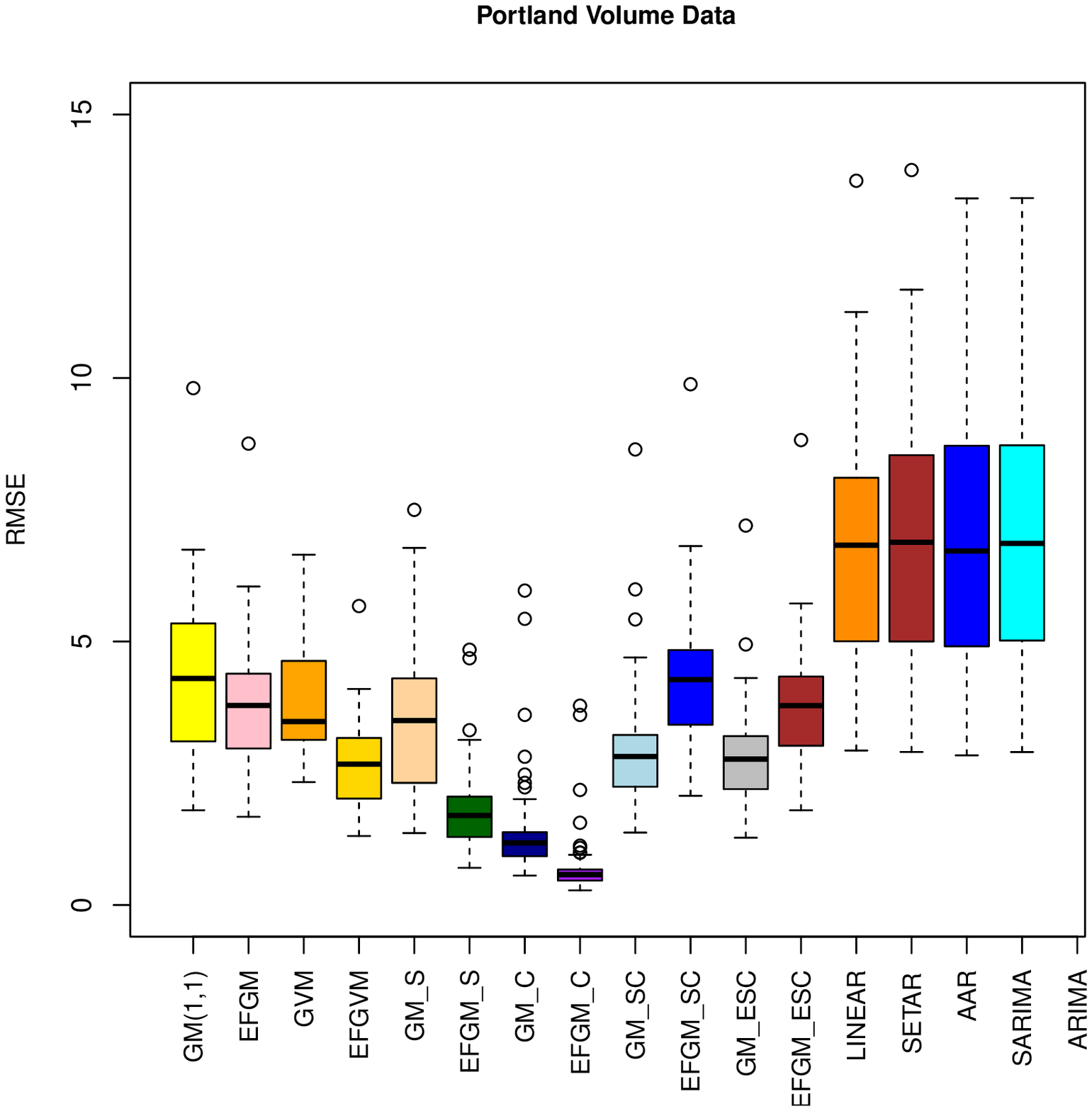}
  \label{fig_rmset}
\end{subfigure}
\caption{RMSEs for $1$-step volume predictions}
\label{fig_boxplot2}
\end{figure*}

The average RMSE and MAPE for the predicted speed (S), travel time (TT), volume (V), and occupancy (O) for the different models are summarized in the Table~\ref{tab_accuracy}. It can be seen that the Grey models outperformed the time series models using  much less training data. Among the GM models, the error corrected by Fourier series GM(1,1$|cos(\omega t)$) model (denoted by EFGM\_C) performed best. The results indicate that the error corrected Grey Verhulst model (EFGVM) is a better choice when data contains high noise and higher nonlinearity.  However, when the noise level is very low and the system is not subject to drastic changes, the model without error correction GM(1,1$|cos(\omega t)$) (GM\_C) can also be suitable. The last row of the table shows percent improvement of simpler trigonometric GM\_C model over best performing nontrigonometric EFGVM model. GM\_C model is able to surpass EFGVM by $59\%$ in RMSE on Portland speed data and $82\%$ in MAPE on Inrix speed and travel time data. The computation time needed to run the model on the INRIX dataset is provided in the last column (CT).  These times are obtained on a PC with a Pentium i5 Quad-Core CPU and 8 GB of memory. The proposed GMs have higher computation times and they can be improved via metaheuristic.
\begin{figure}[h!]
\centering
\begin{subfigure}{0.495\textwidth}
\centering
  \includegraphics[width=0.86\linewidth]{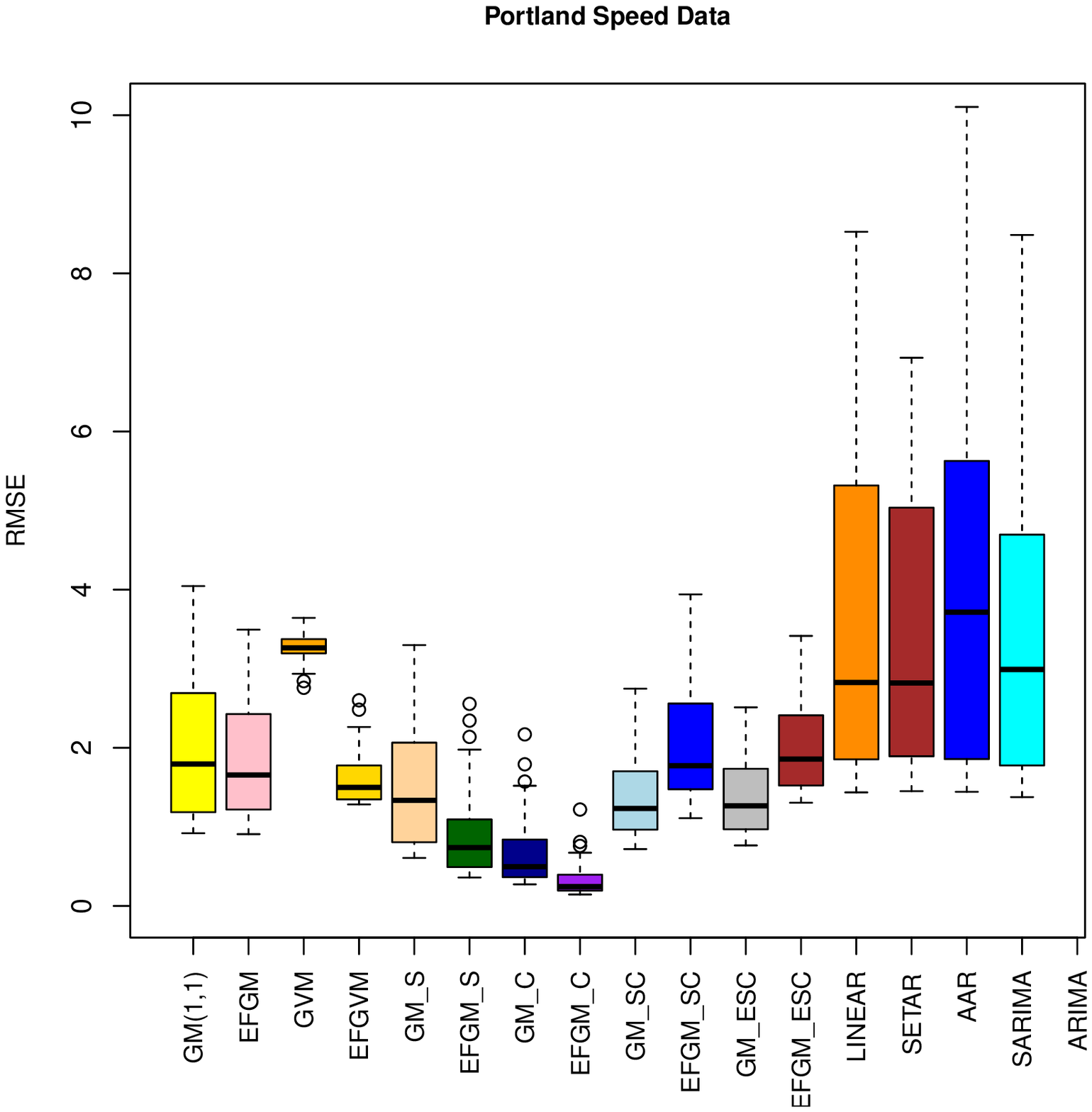}
  \label{fig_boxplot3a}
\end{subfigure}%
\begin{subfigure}{0.495\textwidth}
 \centering
\includegraphics[width=0.86\linewidth]{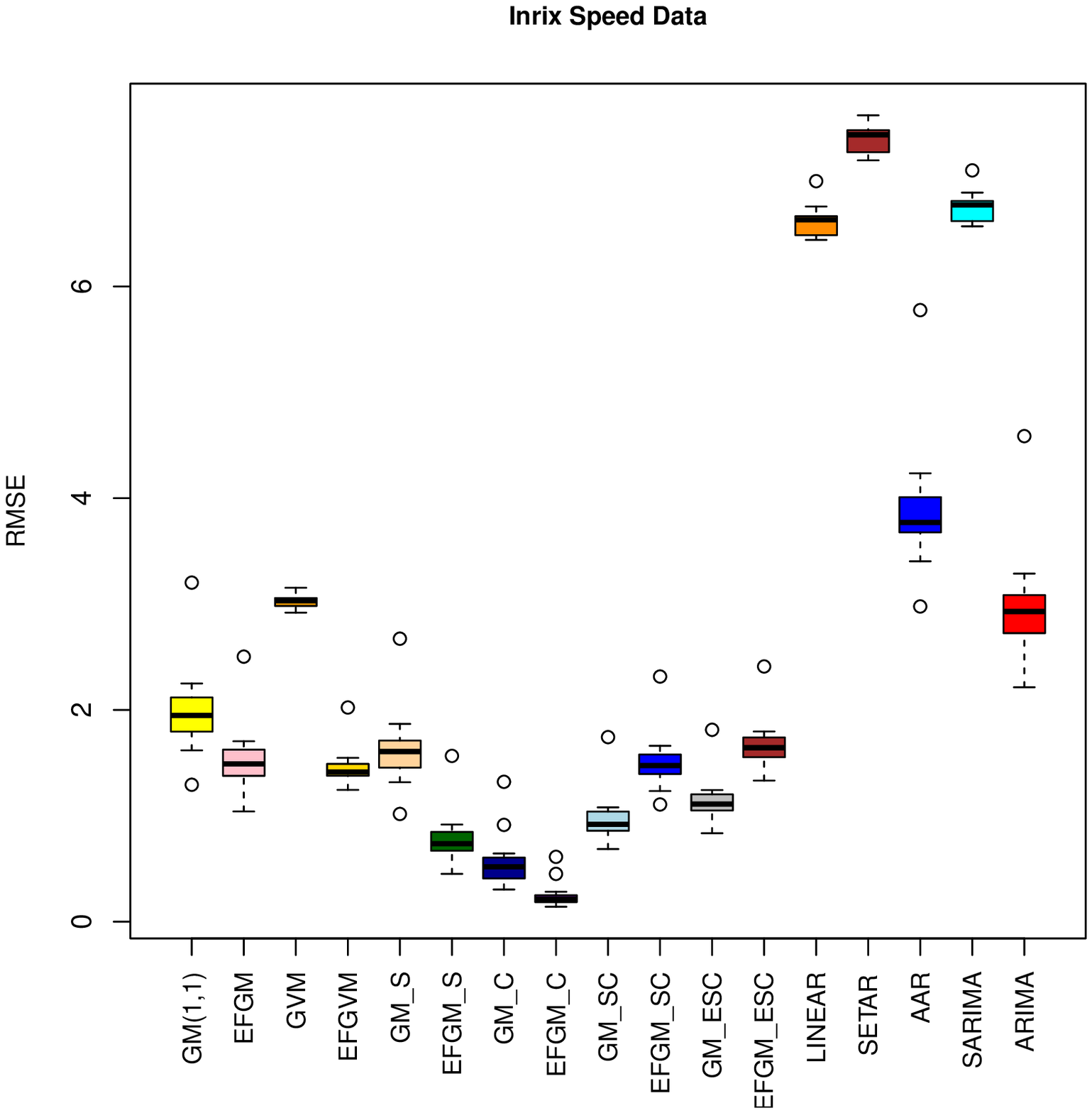}
  \label{fig_boxplot3b}
\end{subfigure}
\caption{RMSEs for $1$-step speed predictions}
\label{fig_boxplot3}
\begin{subfigure}{0.495\textwidth}
\centering
  \includegraphics[width=0.86\linewidth]{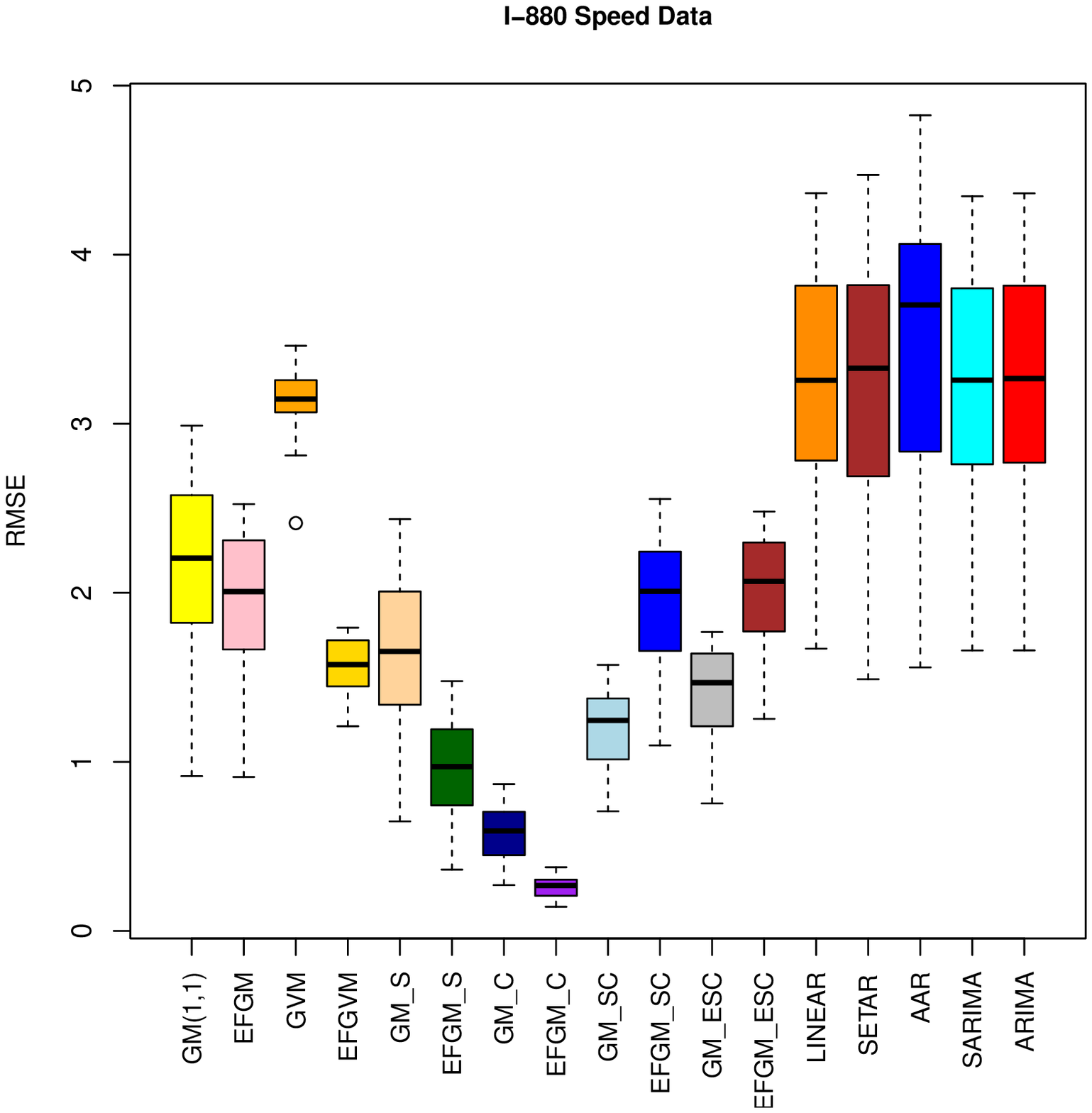}
  \label{fig_boxplot4a}
\end{subfigure}%
\begin{subfigure}{0.495\textwidth}
 \centering
\includegraphics[width=0.86\linewidth]{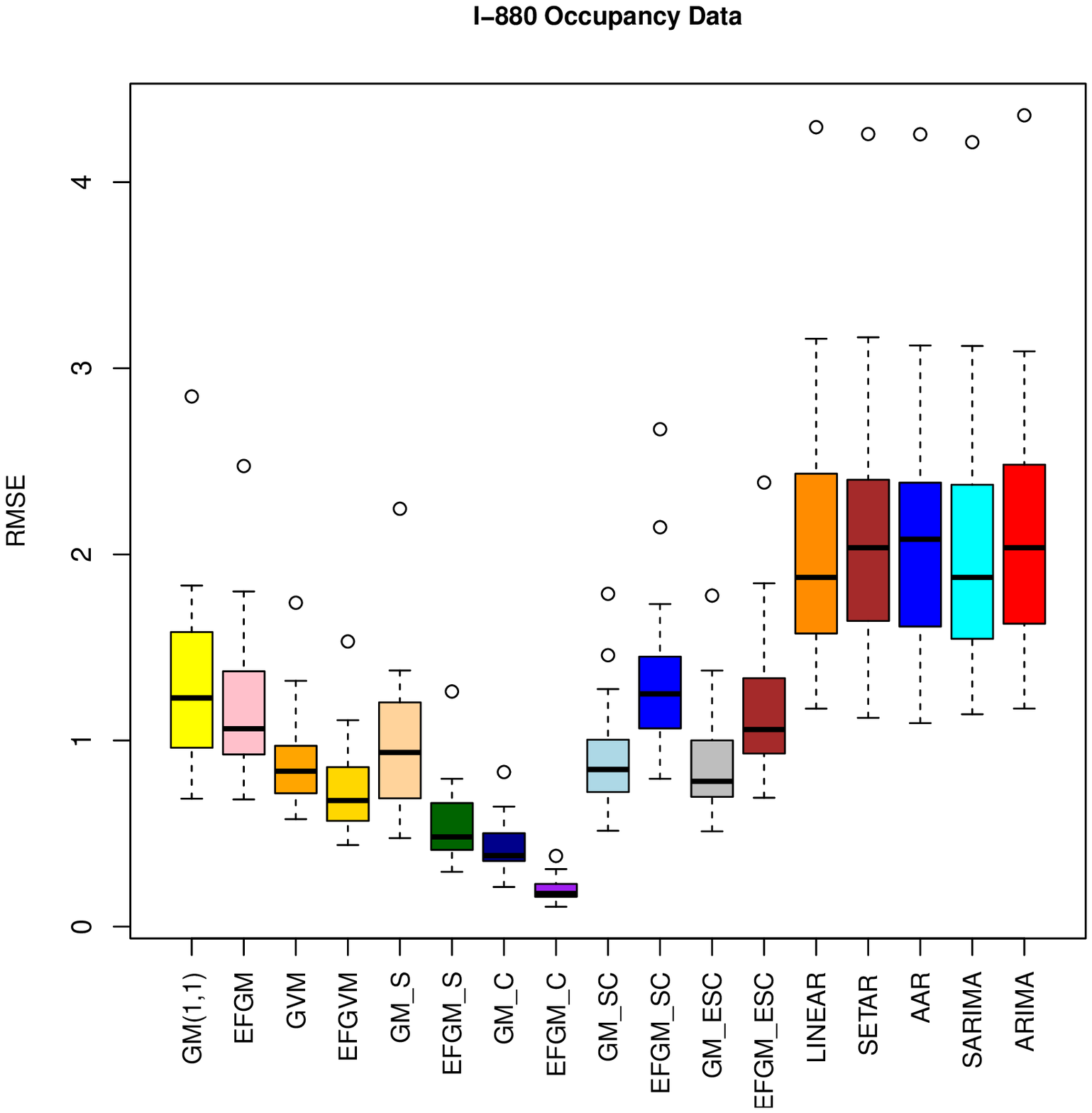}
  \label{fig_boxplot4b}
\end{subfigure}
\caption{RMSEs for $1$-step speed and occupancy predictions}
\label{fig_boxplot4}
\end{figure}

The proposed Grey models are compared to each other to determine which one would produce better results more consistently. Figs.~\ref{fig_boxplot1}-\ref{fig_boxplot4} show RMSEs for each time series and for each model. As shown in the figure, RMSE values are below 3 (mph) for the GM(1,1) model and error corrected GM(1,1) (EFGM); and below 2 (mph) for EFGVM. EFGVM model yielded more accurate predictions for all datasets among models that do not contain trigonometric terms. Further examination shows that EFGM\_C follows the true values closest (see Figs.~\ref{fig_fore1}-\ref{fig_fore2}), especially when there is significant sudden change. The model errors are shown graphically via Box plots in Figs.~\ref{fig_boxplot1}-\ref{fig_boxplot4}. 
\section{Conclusions}
\label{sctconc}
This paper introduced new Grey system models to forecast traffic parameters and compared them to other benchmark models. Overall, the proposed Grey models yielded more accurate forecasts. The GM(1,1$|cos(\omega t)$) model outperformed other Grey models when the system has abrupt changes. The key advantages of using the proposed Grey system models are: (1) simple expert system with well-defined mathematics and straightforward adaptation to include multidimensional data such as weather and other location specific observations, and (2) simple implementation that would allow the model to be implemented and deployed on portable electronic devices such as smart phones and tablets (\cite{vlahogianni2014short}).

Future work in this area could include exploration of whitenization functions, long-term predictions, and the use of multivariable Grey models.  These models could be further extended for use on interrupted facilities (i.e., highways instead of freeways).
\section*{Acknowledgments}
This study is supported by the Center for Connected Multimodal Mobility ($C^2$$M^2$) (USDOT Tier 1 University Transportation Center) Grant headquartered at Clemson University, Clemson, South Carolina, USA. The authors would also like to acknowledge U.S. Department of Homeland Security (DHS) Summer Research Team Program Follow-On, and National Science Foundation (NSF, No. 1719501) grants. Any opinions, findings, conclusions or recommendations expressed in this material are those of the author(s) and do not necessarily reflect the views of ($C^2$$M^2$), USDOT, DHS, or NSF and the U.S. Government assumes no liability for the contents or use thereof.

\bibliographystyle{elsarticle-harv}
\bibliography{detection_trb}
\end{document}